\documentclass[aps,twocolumn,float,prd,psfig,amsmath,amssymb]{revtex4}

\usepackage[dvipdfmx]{graphicx}
\usepackage{dcolumn}
\usepackage{bm}
\usepackage{braket}
\usepackage{braket}
\usepackage{breqn}
\usepackage{color}

\newcommand{\add}[1]{\textcolor{black}{#1}}

\newcommand{\RN}[1]{%
  \textup{\uppercase\expandafter{\romannumeral#1}}%
}

\newcommand{\ro}{\it \RN {1}}
\newcommand{\rt}{\it \RN{2}}

\newcommand{\lsim}{\lesssim}
\newcommand{\gsim}{\gtrsim}

\newcommand{\lmk}{\left(}
\newcommand{\rmk}{\right)}
\newcommand{\lnk}{\left\{ }
\newcommand{\rnk}{\right\} }
\newcommand{\lkk}{\left[} 
\newcommand{\rkk}{\right]}
\newcommand{\lla}{\left\langle}
\newcommand{\p}{\partial}
\newcommand{\rra}{\right\rangle}

\begin{document}


\title{  Correlation of  Gravitational Wave Background Noises and Statistical Loss for Angular Averaged Sensitivity Curves }


\author{Naoki Seto}
\affiliation{Department of Physics$,$ Kyoto University$,$ Kyoto 606-8502$,$ Japan}


\date{\today}

\begin{abstract}

Gravitational wave backgrounds  generate correlated noises to separated detectors.  This correlation can induce statistical losses to actual detector networks,  compared with idealized noise-independent networks. Assuming that the backgrounds are isotropic,  we examine the statistical losses  specifically for the angular averaged sensitivity curves, and derive  simple expressions that depend on the overlap reduction functions and the strength of the  background noises relative to the instrumental noises. For future triangular interferometers such as ET and LISA, we  also discuss  preferred network geometries to suppress the potential statistical losses.


\end{abstract}


\maketitle

\section{Introduction}

 Gravitational wave astronomy has evolved rapidly  after the detection of the first event GW150914 \cite{LIGOScientific:2016aoc,LIGOScientific:2018mvr}.  The sensitivities of the current generation detectors  have been improved gradually \cite{KAGRA:2013pob}.   In addition,  we have various future plans to observe gravitational waves at broad frequency regimes.  For example,  around 10-1000Hz, Einstein telescope (ET) \cite{Hild:2010id}  and Cosmic Explore \cite{Reitze:2019iox} will have $\sim 10$ times better  sensitivities  than advanced LIGO.  Furthermore, they will push the low-frequency noise walls down to $\sim1$Hz.  In space, LISA  \cite{lisa0,lisa}, TianQin \cite{Luo:2015ght,Huang:2020rjf}  and Taiji \cite{taiji} are  proposed to explore the mHz band, while B-DECIGO and DECIGO are targeting the 0.1Hz hand  \cite{Kawamura:2020pcg}.

The scientific prospects of these future projects have been widely discussed with various statistical quantities such as the parameter estimation errors  for individual astrophysical sources and their detectable volumes (see e.g. \cite{KAGRA:2013pob}).  
Here,  some of these quantities depend strongly on the geometries of the detector networks. 

Stochastic gravitational wave backgrounds are  interesting observational targets. For their detection, the correlation analysis  is an effective approach, and enables us to detect a weak background  whose strain spectrum  is  $\sim(fT_{\rm int})^{-1/2}$ times smaller than the instrumental noise spectra (defined in units of $\rm Hz^{-1}$ not $\rm Hz^{-1/2}$) \cite{Christensen:1992wi,Flanagan:1993ix,Allen:1997ad,Romano:2016dpx,LIGOScientific:2019vic}.  Here $f$ is the frequency of the background waves and $T_{\rm int}$ is the integration time for the correlation analysis.

Meanwhile, considering  our limited understandings on high energy physics and early universe, we cannot securely  impose tight upper limits on the magnitudes of  backgrounds, purely from  a theoretical standpoint.  Indeed, there are variety of  cosmological scenarios to generate  backgrounds at various frequency regimes (see e.g. \cite{Caprini:2018mtu}).  Therefore, in the new observational windows opened by the future projects,  practically unconstrained by current observations, we might actually  have a background comparable to the designed  instrumental noises.

The angular averaged sensitivity is one of the basic measures to characterize  gravitational wave detectors \cite{lisa,Cornish:2018dyw}.  If the     noises  of detectors are statistically independent and have an identical spectrum, the angular averaged sensitivity of their network should follow  a simple scaling relation with respect to  the number of detectors.  But, in reality,  gravitational wave backgrounds can induce correlated noises between separated detectors  \cite{Christensen:1992wi,Flanagan:1993ix,Allen:1997ad,LIGOScientific:2019vic}.  Resultantly, the background noises  break the simple scaling relation for the angular averaged sensitivity.  
The situation should depend on the geometry of the detector network and the strength of the background noises relative to the instrumental noises. 

In this paper,   we quantify the statistical losses due to  the background noise correlation, by evaluating the deviation from the simple scaling relation.  We also discuss preferred network geometries to suppress the statistical losses both for ground-based and space-borne detector networks.

  This paper is organized as follows. In Sec. II, we study a basic model for two L-shaped detectors. In Sec. III, we discuss the networks composed by two triangular detectors tangential to a sphere. In Sec. IV,  for future detectors, we numerically  discuss the dependence of the statistical losses on the network geometry.  Sec. V is devoted to summary and discussion. 
  

\section{angular averaged sensitivity}

\subsection{Noise Components}
We consider an effectively L-shaped interferometer  (with the label $\ro$)  and decompose its data stream $a_\ro$ in the Fourier space as
\begin{eqnarray}
a_\ro(f)=h_\ro(f)+n_{\ro D}(f)+n_{\ro B}(f).
\end{eqnarray}
Here $h_\ro(f)$ represents a  gravitational wave signal (e.g. from a compact binary), $n_{\ro D}(f)$    the instrumental noise and $n_{\ro B}(f)$  the  noise due to isotropic gravitational wave backgrounds.  The two noises are assumed to be stationary and Gaussian distributed.  We define the instrumental noise  spectrum $N_{D}(f)$ by 
\begin{eqnarray}
\lla n_{\ro D}(f)n_{\ro D}(f')^*\rra=\delta(f-f') N_D(f)
\end{eqnarray}
with the ensemble average $\lla \cdots\rra$ and the delta function $\delta(\cdot)$.  We also  define the background noise spectrum $N_B(f)$ by 
\begin{eqnarray}
\lla n_{\ro B}(f)n_{\ro B}(f')^*\rra=\delta(f-f') N_B(f).
\end{eqnarray}
Hereafter, for notional simplicity, we omit the delta functions, using expressions for $f=f'$.

Similarly, we consider the second L-shaped interferometer $\rt$ and  represent its  data  stream by
\begin{eqnarray}
a_{\rt}(f)=h_{\rt}(f)+n_{\rt D}(f)+n_{\rt B}(f).
\end{eqnarray}
We assume that  its  instrumental noise  spectrum is identical to the first one $\ro$ as 
\begin{eqnarray}
\lla n_{\rt D}(f)n_{\rt D}(f)^*\rra= N_D(f) ,
\end{eqnarray}
but is statistically independent as
$\lla n_{\ro D}(f)n_{\rt D}(f)^*\rra=0$.  Actually,  a weak correlation of instrumental noises does not largely change  the present study.  This is  different from the requirement at detecting a weak gravitational wave background by the correlation analysis (see e.g. \cite{Thrane:2013npa,Kowalska-Leszczynska:2016low,Himemoto:2019iwd}).

For the background noise of the second interferometer $\rt$, we put
\begin{eqnarray}
\lla n_{\rt B}(f)n_{\rt B}(f)^*\rra= N_B(f) .
\end{eqnarray}
We should notice that,  in contrast to the  instrumental noises, the background noises would  have a definite correlation 
\begin{eqnarray}
\lla n_{\ro B}(f)n_{\rt B}(f)^*\rra= N_B(f) \gamma(f)\label{orf1}
\end{eqnarray}
characterized by the overlap reduction function (ORF) $\gamma(f)$ with $-1\le \gamma (f) \le 1$ (see e.g. \cite{Christensen:1992wi,Flanagan:1993ix,Allen:1997ad,Matas:2020roi}).

Then, the noise matrix can be expanded as ($i,j=\ro,\rt$)
\begin{eqnarray} 
N_{ij}(f)&\equiv&  \lla  \lnk  n_{i D}(f)+ n_{i B}(f)  \rnk^*  \lnk  n_{j D}(f)+ n_{j B}(f)  \rnk  \rra\nonumber \\
&=&
(N_D(f)+N_B(f))\begin{pmatrix}
1 & \frac{K\gamma}{(1+K)} \\
\frac{K\gamma}{(1+K)} & 1 \\
\end{pmatrix}.\label{nm}
\end{eqnarray}
Here we defined the ratio between the two noise components  by
\begin{eqnarray}
K(f)\equiv \frac{N_B(f)}{N_D(f)}.
\end{eqnarray}
 The two parameters $\gamma$ and $K$ play important roles in the analysis below.

\subsection{Averaged Signal-to-Noise Ratio}
Next, we discuss the angular averaged sensitivities of detector networks to gravitational wave signals.  To begin with, we examine the signal analysis with the single interferometer $\ro$, and put its gravitational wave signal $h_\ro(f,\alpha)$. Here we introduced the parameter $\alpha$ to abstractly show the polarization and direction angles.   As in the case of evaluating the angular averaged sensitivity curve, we take the following  ratio as an intermediate product \cite{Cutler:1994ys} 
\begin{eqnarray}
\sigma_1=\frac{\lla h_\ro(f,\alpha)  h_\ro(f,\alpha)^*  \rra_{\alpha}}{N_D(f)+N_B(f)},\label{sin}
\end{eqnarray}
where $\lla \cdots\rra_\alpha$ represents the averages with respect to the polarization and direction angles. 
\add{Roughly speaking, the square root of $\sigma_1$ is proportional to the signal strength relative to the noise around the frequency $f$.  Since we will  soon compare $\sigma_1$ with $\sigma_{\cal N}$ defined for multiple detectors, the common factors were dropped in Eq. (\ref{sin}).  }

For a network composed by totally $\cal N$ detectors of the identical specifications, we can extend Eq. (\ref{sin}) in the matrix form as \cite{Cutler:1994ys} 
\begin{eqnarray}\label{mat0}
\sigma_{\cal N}=\sum_{ij}^{\cal N}{\lla h_i(f,\alpha)  h_j(f,\alpha)^*  \rra_{\alpha}}{N_{ij}(f)^{-1}}.
\end{eqnarray}
Then we use the ratio $\sigma_{\cal N}/\sigma_1$ to measure  the statistical  gain of the  angular averaged  sensitivity by using $\cal N$ detectors,  compared with the single one. 
For a network with noise independent equivalent detectors, we readily obtain $\sigma_{\cal N}/\sigma_1={\cal N}$.
  On the other hand, if  we have correlated background noises, the effective number of detectors could be smaller than $\cal N$.  
  

Next, we specifically examine the case ${\cal N}=2$. 
For the signal matrix $\lla  h_i(f,\alpha)  h_j(f,\alpha)\rra_\alpha$  (with $i,j=\ro,\rt$),  the diagonal elements have the relation
\begin{eqnarray}
\lla h_\ro(f,\alpha)  h_{\ro }(f,\alpha)^*  \rra_{\alpha}=\lla h_{\rt}(f,\alpha)  h_{\rt}(f,\alpha)^*  \rra_{\alpha}.
\end{eqnarray}
The off-diagonal elements can be expressed as \cite{Christensen:1992wi,Flanagan:1993ix,Allen:1997ad,Romano:2016dpx}
\begin{eqnarray}
 \lla h_\ro(f,\alpha)  h_{\rt}(f,\alpha)^*  \rra_{\alpha}&=&\lla h_{\rt}(f,\alpha)  h_{\ro }(f,\alpha)^*  \rra_{\alpha}\\
&=& \gamma(f) \lla h_\ro(f,\alpha)  h_{\ro }(f,\alpha)^*  \rra_{\alpha}
\end{eqnarray}
with the ORF $\gamma$ already used in Eq. (\ref{orf1}).

Now, we evaluate the statistical gain  $\sigma_2/\sigma_1$.  
As mentioned earlier, this shows the  improvement of the angular averaged sensitivity by using the two detectors. 
Appropriately cancelling common factors, we have 
\begin{eqnarray}
\frac{\sigma_2}{\sigma_1}&=&{\rm Tr} \lkk 
\begin{pmatrix}
1 & \gamma \\
\gamma & 1 \\
\end{pmatrix}
\begin{pmatrix}
1 & \frac{K\gamma}{(1+K)} \\
\frac{K\gamma}{(1+K)} & 1 \\
\end{pmatrix}^{-1}\rkk \label{matr} \\
&=&\frac{2(1+K)(1+K-\gamma^2K)}{(1+K)^2- K^2 \gamma^2}\label{defg}\\
&\equiv&  G_2(K,\gamma).
\end{eqnarray}
In Fig. 1, we present a contour plot for the analytic  expression $ G_2(K,\gamma)$ which is an even function of $\gamma$.

We can easily confirm that the maximum  value of $G_2$ is realized at  $K=0$ or $\gamma=0$ as 
\begin{eqnarray}
 G_2(0,\gamma)= G_2(K,0)=2.
\end{eqnarray}
For these sets of parameters $(\gamma,K)$, the noise matrix (\ref{nm}) becomes diagonal and the two detectors are statistically independent.  We thus obtain $G_2=2$, as easily expected.  \add{ The deficit $2-G_2(K,\gamma)$ shows the statistical loss induced by the background  noise correlation.}

For $K\ll 1$, we can expand $G_2(K,\gamma)$ as
\begin{eqnarray}
G_2(K,\gamma)=2(1-K\gamma^2)+O(K^2).\label{gsk}
\end{eqnarray}

In the limit $K\to \infty$, we have
\begin{eqnarray}
\lim_{K\to \infty} G_2(K,\gamma)=2 \label{lim1}
\end{eqnarray}
for $\gamma\ne\pm1$.  In contrast, we obtain
\begin{eqnarray}
\lim_{K\to \infty} G_2(K,\gamma=\pm1)=1.
\end{eqnarray}
Therefore, we have the inequality 
\begin{eqnarray}
\lim_{\gamma\to\pm1}[\lim_{K\to \infty} G_2(K,\gamma)]\ne\lim_{K\to \infty}[\lim_{\gamma\to \pm1} G_2(K,\gamma)],
\end{eqnarray}
 showing the dependence on the order of the limiting operations.  

At first sight, Eq. (\ref{lim1}) might look counterintuitive. For example, let us consider two highly correlated detectors with $\gamma=0.999$.  They have the nearly the same signals
\begin{eqnarray}
h_\ro(f,\alpha)\sim h_\rt(f,\alpha)
\end{eqnarray}
and, at $K\gg 1$, their noises also satisfy
\begin{eqnarray}
n_{\ro D}(f)+n_{\ro B}(f)\sim n_{\rt D}(f)+n_{\rt B}(f)
\end{eqnarray}
with the relation $n_{\ro B}(f)\sim n_{\rt B}(f)$ for $\gamma\sim1$.  But this does not result in ${G_2}\sim 1$. 
 In fact, we can diagonalize the noise matrix (\ref{nm}) by taking the linear combination of the original  data streams as 
\begin{eqnarray}
a'=\frac{(a_\ro-a_{\rt})}{\sqrt2},~~a''=  \frac{(a_\ro+a_{\rt})}{ \sqrt2}.
\end{eqnarray}  
At $K\to \infty$, the differential mode $a'$ simultaneously reduces the largely overlapped background noises and the gravitational wave signals. 
More specifically, the noise spectrum of the differential mode $a'$ is given by
\begin{eqnarray}
N_D+N_B(1-\gamma) \label{na1}
\end{eqnarray}
and the signal strength by
\begin{eqnarray}
\lla h_\ro(f,\alpha) h_\ro(f,\alpha)^*\rra_\alpha (1-\gamma).
\end{eqnarray}
Meanwhile  the total mode $a''$ has   the noise spectrum
\begin{eqnarray}
N_D+N_B(1+\gamma)
\end{eqnarray}
and the signal
\begin{eqnarray}
\lla h_\ro(f,\alpha) h_\ro(f,\alpha)^*\rra_\alpha (1+\gamma).
\end{eqnarray}
Thus, even for $\gamma\sim 1$ (but $\gamma \ne 1$),     the  signal-to-noise ratio of the differential mode $a'$ is comparable to the total mode  $a''$ at $K=N_B/N_D\to \infty$.   In contrast, at $K\sim (1-\gamma)^{-1}$, this comparability is  destroyed by the incoherent instrumental noise contribution $N_D(f)$ in Eq. (\ref{na1}).  

For a newly opened frequency window of gravitational waves, the ratio $K$ cannot be securely predicted ahead of time.  In principle, we can make  theoretical estimations based on some models, or  use  cosmological constraints on primordial waves (see e.g. \cite{Smith:2006nka}).  But it would be fruitful to extract insights that are independent of the ratio $K$.  To this end, we evaluate the minimum value of $G_2(K,\gamma)$ for a given $\gamma$, as the worst case.  By solving the equation $\p_K G_2=0$ for $K$, we find  the solution $K_{\rm min}(\gamma)=(1-\gamma^2)^{-1/2}$ and obtain the corresponding minimum $G_2[K_{\rm min}(\gamma),\gamma]= 1+\sqrt{1-\gamma^2}\equiv G_{2\rm \min}(\gamma)$ with
\begin{equation}
 G_2(K,\gamma)\ge G_{2\rm \min}(\gamma).\label{min}
\end{equation}

\add{For the number of detectors  ${\cal N}>2$,  Eq.  (\ref{lim1}) is generalize to  $\lim_{K\to \infty}\sigma_{\cal N}/\sigma_1={\cal N}$, if none of the overlap reduction functions are 1 or $-1$.  This is because the instrumental noises can be ignored and Eq. (\ref{matr})  becomes the trace of the $\cal N$-dimensional  unit matrix. 
}

\begin{figure}[t]
\centering
\includegraphics[keepaspectratio, scale=0.6]{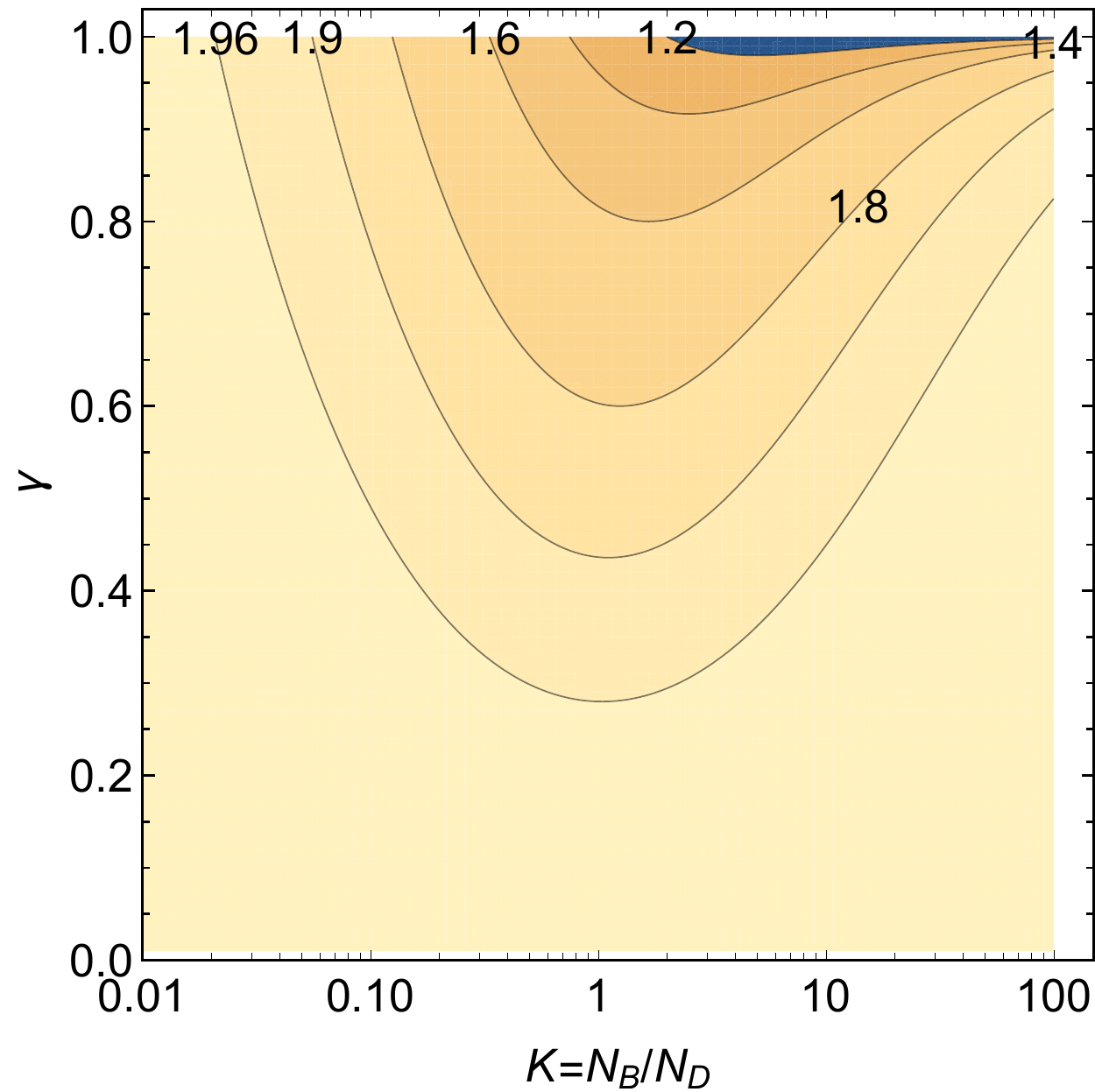}
\caption{Contour plot for the effective detector number $ G_2(K,\gamma)$ defined for two L-shaped interferometers with the same instrumental noise spectrum. Their background noises have correlation, characterized by the ORF $\gamma$.  The parameter $K$ represents the ratio between the background and instrumental noises.     We have $1\le G_2(K,\gamma)\le 2$ with $ G_2(0,\gamma)= G_2(K,0)=2$. 
}
\label{fig:1}
\end{figure}

\section{Two triangular detectors on a sphere}

 The ground based detector ET \cite{Hild:2010id} and the space-borne detector LISA \cite{lisa} are both planned to have triangular geometry.   Such detectors have preferable symmetric structure and enable us to simplify otherwise cumbersome calculations.   Therefore, below, we focus on networks composed by two equivalent  triangular units.   We should mention that the geometrical arguments in this section share similarities with Ref. \cite{Seto:2020zxw,Omiya:2020fvw}  done for a largely different topic  (polarization modes of gravitational wave backgrounds). 

\subsection{Data Streams From a Single Triangular Unit}

First, we briefly discuss the internal symmetry of a triangular unit.  We assume that we can make three effective interferometers $X,Y$ and $Z$ symmetrically at each vertex, as shown in Fig. 2.

\begin{figure}[t]
\centering
\includegraphics[keepaspectratio, scale=0.4]{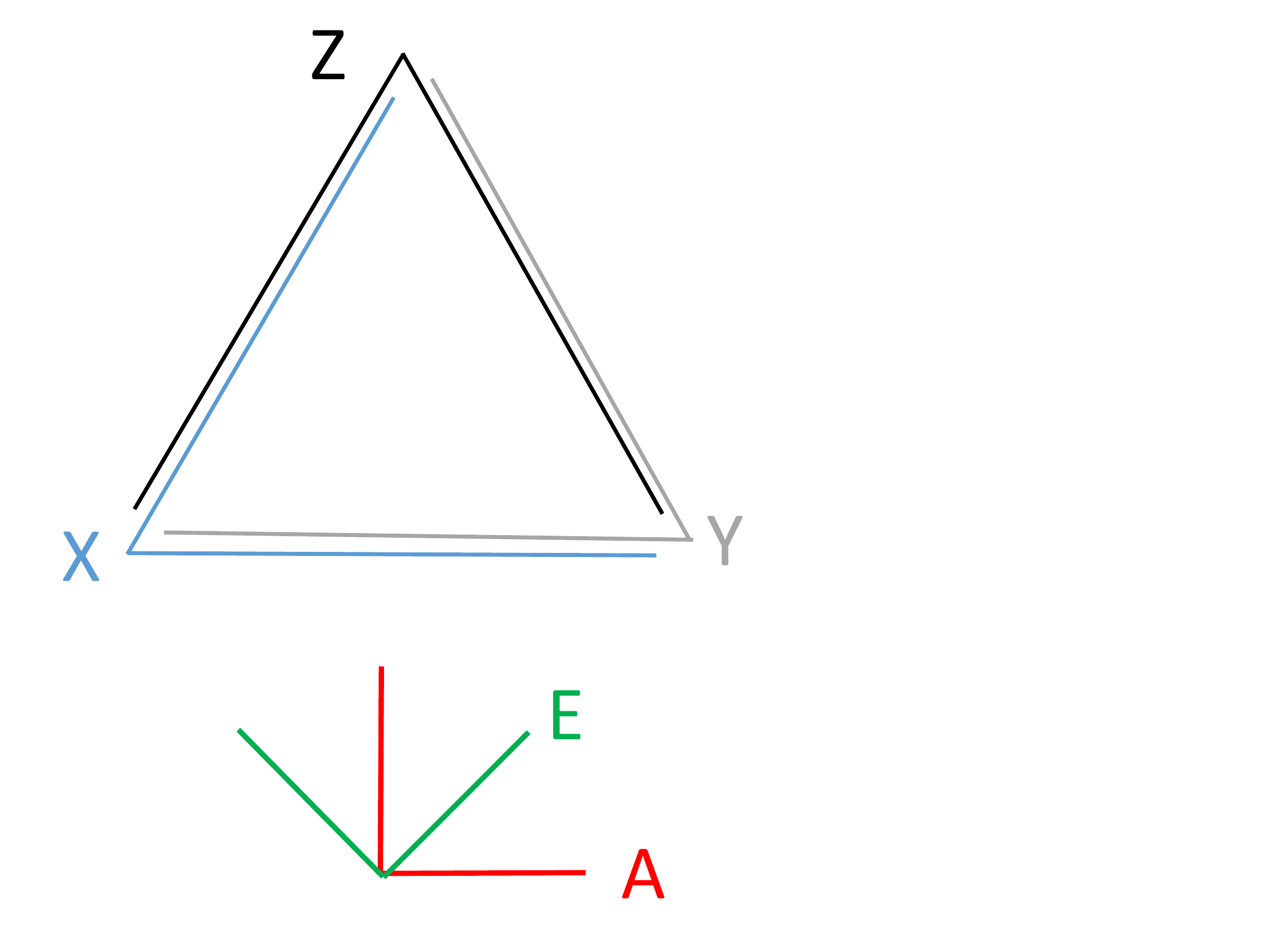}
\caption{The orientations of  the two effective interferometers  $A$ and $E$ made from the symmetric data streams $X,Y$ and $Z$.   The instrumental noises of $A$ and $E$ have no correlation, and their ORF is   $\gamma_{AE}=0$.
}
\end{figure}

Next, we study the $3\times3$ matrix for their instrumental noises $(n_X,n_Y,n_Z)$, omitting the apparent frequency dependence, for notational simplicity.  Given the symmetry of the system, we can put
\begin{eqnarray}
\lla n_X n_X^*\rra=\lla n_Y n_Y^*\rra=\lla n_Z n_Z^*\rra=N_{d}
\end{eqnarray}
for the diagonal elements of the instrumental noise matrix.   For the six off-diagonal elements, considering the symmetry and the potential correlations (e.g. due to the seismic noises for ET), we put 
\begin{eqnarray}
\lla n_X n_Y^*\rra=\lla n_Y n_X^*\rra=\lla n_X n_Z^*\rra=\cdots=N_{o}.
\end{eqnarray}
Therefore, the instrumental noises matrix becomes symmetric and can be diagonalized with an orthogonal matrix.  For example, we take the three data combinations \cite{Prince:2002hp,Mentasti:2020yyd}
\begin{eqnarray}
A&=&\frac{X-Y}{\sqrt2}, ~~E=\frac{X+Y-2Z}{\sqrt6},\label{dae}\\
T&=&\frac{X+Y+Z}{\sqrt3}\label{dt}.
\end{eqnarray}
Their instrumental noise matrix has the following diagonal components
\begin{eqnarray}
N_{AA}&=&N_{EE}=N_{d}-N_{o},\label{nae}\\
N_{TT}&=&N_{d}+2N_{o}
\end{eqnarray}
with the off-diagonal ones $N_{AE}=N_{ET}=N_{TA}=\cdots=0$.
Similarly using the underlying symmetry, we can also confirm the following relations for the ORFs
\begin{eqnarray}\label{gae}
\gamma_{AE}=\gamma_{ET}=\gamma_{TA}=0.
\end{eqnarray}

The information content is the same for $(X,Y,Z)$ and $(A,E,T)$.   But the latter would be more advantageous for the present study, exploiting  the symmetries of the system (including the case with $N_o(f)=0$). 

In the next section, we discuss the statistical loss induced by the background noise correlation for two triangular units.  Such effect is more important in the lower frequency regime.  But, there,  the sensitivity of the $T$-mode is much worse than the $A$ and $E$ modes, due to a signal cancellation.  We can easily confirm this cancellation  by applying the low frequency approximation to the original data  $(X,Y,Z)$ and then evaluating Eqs. (\ref{dae}) and (\ref{dt}).
Below, we only keep the $A$ and $E$ modes that can be regarded as the two L-shaped interferometers whose orientations are shown in Fig. 2.  In the connection to the previous section, we can put $N_D=N_d-N_o$ for their instrumental noise spectrum.

In fact,  as shown in Eq. (\ref{nae}),  the eigen values for the noise matrix are degenerated for the  $A$ and $E$ modes, and we have the freedom to additionally introduce an orthogonal matrix and generate the new data combination  \cite{Seto:2020zxw}
\begin{eqnarray}
A(\phi)&=&A\cos2\phi+E\sin2\phi,\label{ar1}\\
 E(\phi)&=&-A\sin2\phi+E\cos2\phi,\label{ar2}
\end{eqnarray}
keeping Eqs. (\ref{nae}) and (\ref{gae}) invariant.  As shown in the bottom panel of  Fig. 2, this arrangement corresponds to virtually  rotating the two interferometers $A$ and $E$ counterclockwise by the angle $\phi$.  The factor of 2 is due to the spin-2 nature of the detector tensor.

\subsection{Two Triangular Units on a Sphere}

Next, we discuss two triangular units $U_1$ and $U_2$ that have identical specifications and are tangential to a sphere with radius $R$.  We put $(A,E)$ for the data streams of $U_1$ and $(A',E')$ for $U_2$.  Our primary objective in this subsection is to evaluate  effective gain $\sigma_{2U}/\sigma_{1U}$ by using the two units $U_1+U_2$ compared with the single unit $U_1$.

\begin{figure}[t]
\centering
\includegraphics[keepaspectratio, scale=0.4]{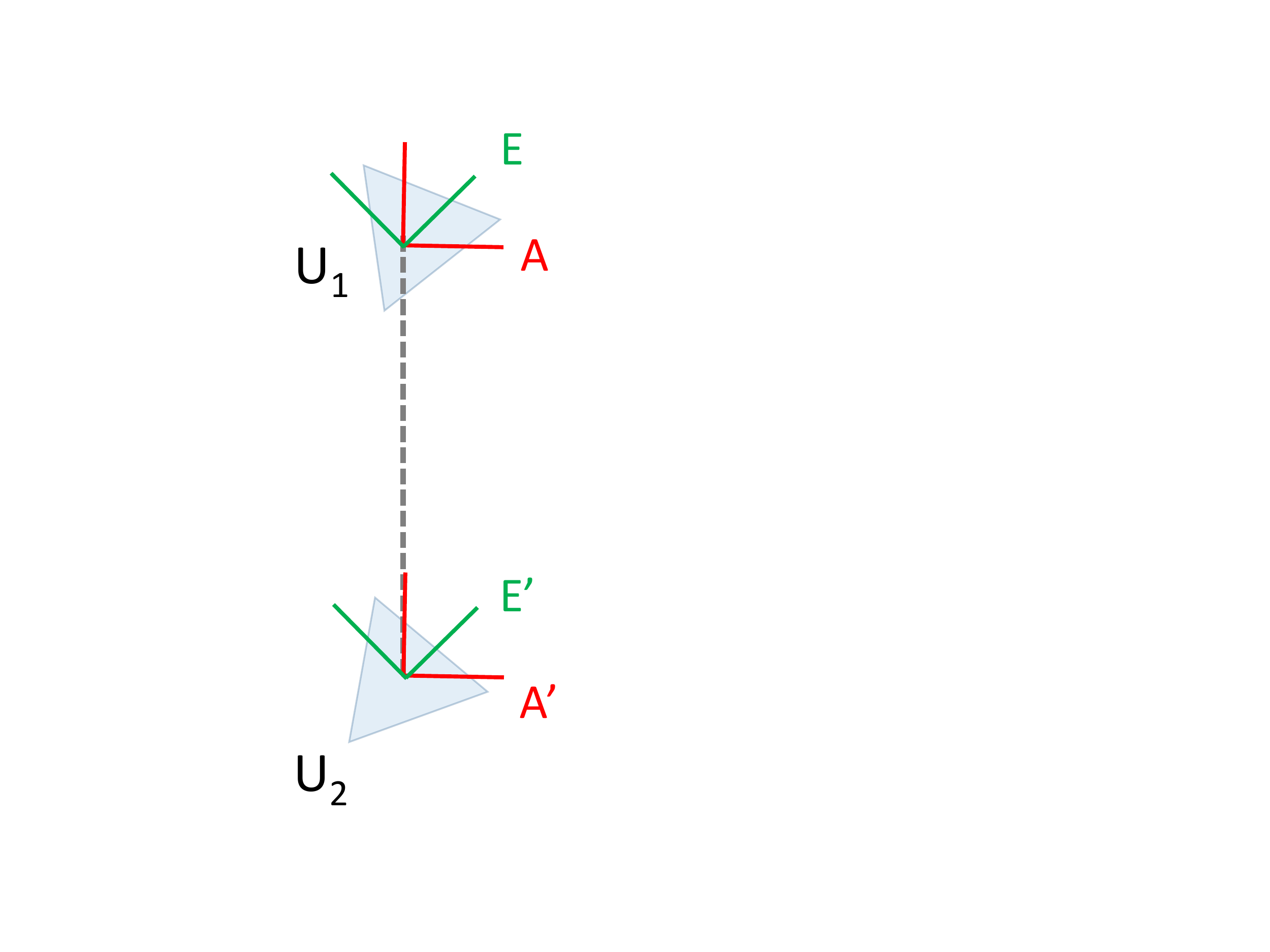}
\caption{Adjusted orientations of the effectively L-shaped interferometers $A,E,A'$ and $E'$.  These four are  generated from the two triangular detectors $U_1$ and $U_2$  tangential to  a sphere. The dashed line shows the geodesic (great circle) connecting the two triangle on the sphere. The effective interferometers $A$ and $A'$ have arms parallel or perpendicular to the geodesic, and $E$ and $E'$ are misaligned  by $45^\circ$.  Due to the geometrical symmetry, we have the ORFs $\gamma_{AE}=\gamma_{A'E'}=\gamma_{AE'}=\gamma_{A'E}=0$.   
}
\end{figure}

Since we have the four effective interferometers $(A,E,A',E')$, the matrices in Eq. (\ref{mat0})  would be $4\times 4$.  However, applying a simple trick associated with Eqs. (\ref{ar1}) and (\ref{ar2}), we can largely simplify the problem. 
As pointed out in \cite{Seto:2020zxw,Omiya:2020fvw}, by using the freedom of the virtual rotation, we can align the orientations of the effective interferometers  $(A,E,A',E')$, following the great circle connecting the two units on the contact sphere (see Fig. 3). 
Below, we use the notations   $(A,E,A',E')$  to represent  the interferometers after the adjustments. 

Using the reflection symmetry of the system, we can show 
\begin{eqnarray}
\gamma_{AE'}=\gamma_{EA'}=0
\end{eqnarray}
along with $\gamma_{AE}=\gamma_{A'E'}=0$ (see Eq. (\ref{gae})) \cite{Seto:2020zxw}.  Then, the $4\times 4$ matrices are block diagonalized for the pairs $(A,A')$ and $(E,E')$.  Accordingly, the effective number of the triangular units  $\sigma_{2U}/\sigma_{1U}$  can be simply expressed by
\begin{eqnarray}\label{gd}
 G_{\Delta}(K,\gamma_{AA'},\gamma_{EE'})= \frac12   \lnk G_2(K,\gamma_{AA'})+G_2(K,\gamma_{EE'})\rnk .
\end{eqnarray}
\add{A single triangular unit provided the two independent interferometers $A$ and $E$ with $\gamma_{AE}=0$. We thus have $\sigma_{1U}/\sigma_1=2$ with $\sigma_1$ defined for the $A$ (or $E$) interferometer alone.  Then we have $\sigma_{2U}/\sigma_{1U}=\sigma_1/\sigma_{1U}\cdot \sigma_{2U}/\sigma_1$  as in Eq. (\ref{gd}).}

From Eq. (\ref{defg}), the expression (\ref{gd}) does not depend on the signs of $\gamma_{AA'}$ and $\gamma_{EE'}$.  
Note that our method based on the symmetry cannot be applied to networks with more than two triangular  units.
\add{From Sec. III, the maximum value of the statistical  gain is  $G_\Delta=2$ (e.g. at $K=0$).  Thus, the statistical loss induced by the background  noise correlation is estimated to be $2-G_\Delta$. }

For $K\ll 1$, from Eq. (\ref{gsk}),  we have
\begin{eqnarray}\label{gd2}
G_{\Delta}(K,\gamma_{AA'},\gamma_{EE'})=2-K(\gamma_{AA'}^2+\gamma_{EE'}^2)+O(K^2).
\end{eqnarray}
This expression might be useful for a  weak background noise.  
Given $|\gamma_{AA'}|\le 1$ and  $|\gamma_{EE'}|\le 1$,  the loss can be at most  $\sim 2K$ for $K\ll 1$. 

\subsection{Minimum Values}
As the worst case similar to   Eq. (\ref{min}), we can evaluate the minimum value  $G_{\Delta \min}(\gamma_{AA'},\gamma_{EE'})$  of the function $G_{\Delta}(K,\gamma_{AA'},\gamma_{EE'})$ by changing $K$ for given $\gamma_{AA'}$ and $\gamma_{EE'}$.  The corresponding point $K=K_{\min}(\gamma_{AA'},\gamma_{EE'})$ is given as a solution of  a hextic polynomial equation $\p_KG_\Delta=0$ whose coefficients are given  by $\gamma_{AA'}$ and $\gamma_{EE'}$.   Then we can formally express the  minimum value by
\begin{eqnarray}
G_{\Delta \min}(\gamma_{AA'},\gamma_{EE'})&\equiv& 
 G_{\Delta}[K_{\min}(\gamma_{AA'},\gamma_{EE'}),\gamma_{AA'},\gamma_{EE'}].\nonumber\\
 & & \label{min1}
\end{eqnarray}

\begin{figure}[t]
\centering
\includegraphics[keepaspectratio, scale=0.6]{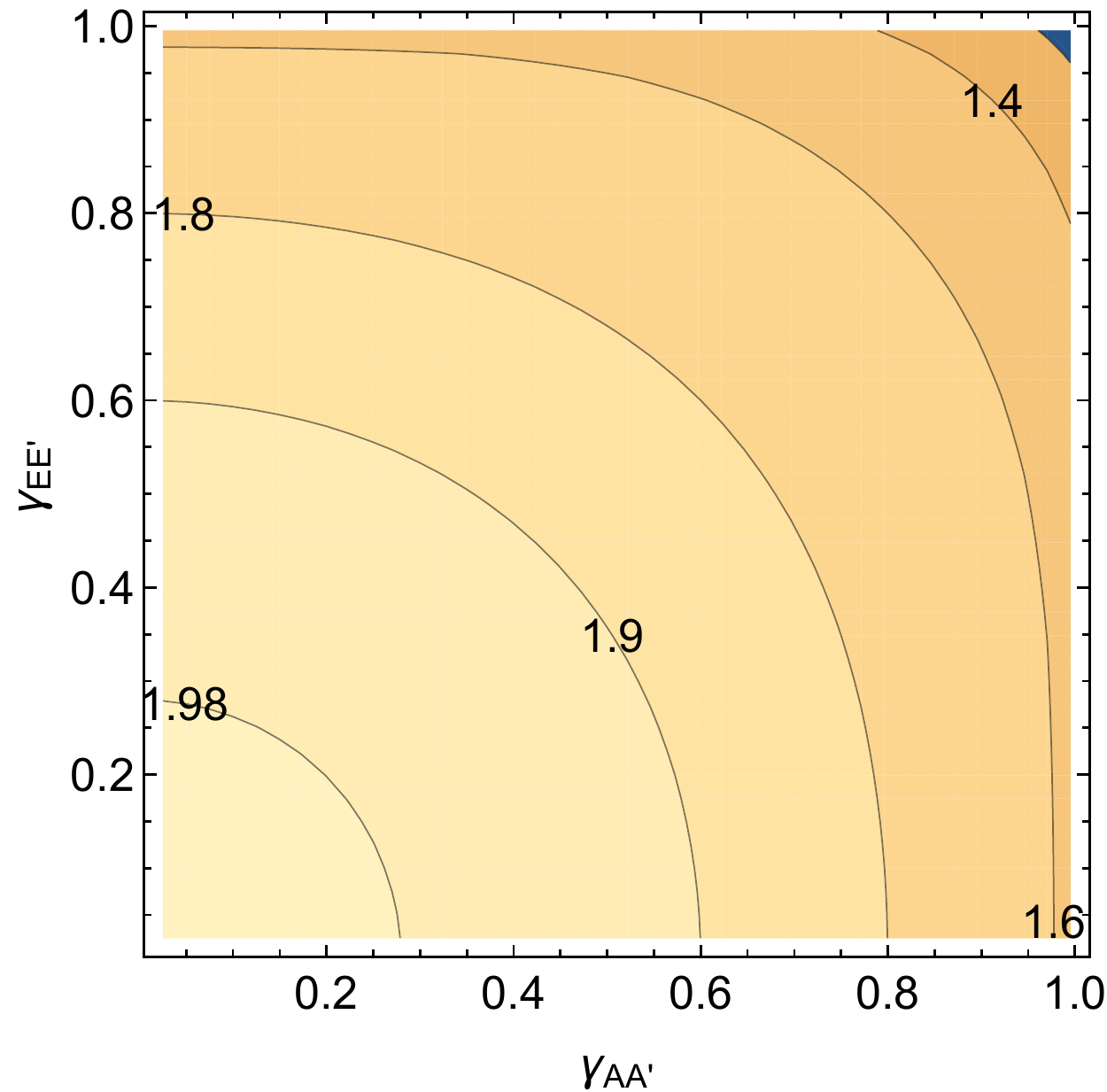}
\caption{Contour plot for the minimum value $G_{\Delta \min}(\gamma_{AA'},\gamma_{EE'})$ of the statistical gain showing the effective number of triangular units.  This function is symmetric with respect to the two arguments $\gamma_{AA'}$ and $\gamma_{EE'}$.  We have  $G_{\Delta \min}(0,0)=2$ and $G_{\Delta \min}(1,1)=1$.
}
\end{figure}

On the other hand, by applying inequality (\ref{min}) individually to the two terms in the right hand  side of Eq. (\ref{gd}), we can also obtain a weaker  bound as
\begin{eqnarray}
G_{\Delta \min}^\dag(\gamma_{AA'},\gamma_{EE'})&\equiv& \frac12 \lkk  G_{2\min}(\gamma_{AA'})+G_{2\min}(\gamma_{EE'}) \rkk \nonumber\\
& & 
\end{eqnarray}
with the function $G_{2\min}(x)\equiv 1+\sqrt{1-x^2}$ defined for  Eq. (\ref{min}).  The two functions $G_{\Delta \min}$ and  $G^\dag_{\Delta \min}$  are symmetric with respect to the arguments $(\gamma_{AA'},\gamma_{EE'})$.
 From their definitions, we have
 \begin{eqnarray}
 G_{\Delta}(K,\gamma_{AA'},\gamma_{EE'})&\ge& G_{\Delta \min}(\gamma_{AA'},\gamma_{EE'})\nonumber\\
& \ge& G_{\Delta \min}^\dag(\gamma_{AA'},\gamma_{EE'}).
\end{eqnarray}
In fact,  the simple function $G^\dag_{\Delta \min}$ is  a good approximation to the complicated one $G_{\Delta \min}$.  
We numerically examined the difference 
\begin{eqnarray}
 G_{\Delta \min}(\gamma_{AA'},\gamma_{EE'})- G_{\Delta \min}^\dag(\gamma_{AA'},\gamma_{EE'})
\end{eqnarray}
in the two dimensional region $0\le  \gamma_{AA'} \le  \gamma_{EE'}\le 1$.   It   exactly vanishes on the two boundaries:  $\gamma_{AA'}=\gamma_{EE'}$ and    $\gamma_{AA'}=0$. 
The  maximum value of the difference is $\sim 0.08$ around $(\gamma_{AA'},\gamma_{EE'})=(0.7,1)$.   But the difference  is
less than 0.01 for the slightly restricted  region
\begin{eqnarray}
0\le \gamma_{AA'}\le \gamma_{EE'}\le 0.93.
\end{eqnarray}
Below, we numerically handle the full expression $ G_{\Delta \min}$ without using its approximation  $G^\dag_{\Delta \min}$.

\begin{figure}[]
\centering
\includegraphics[keepaspectratio, scale=0.6]{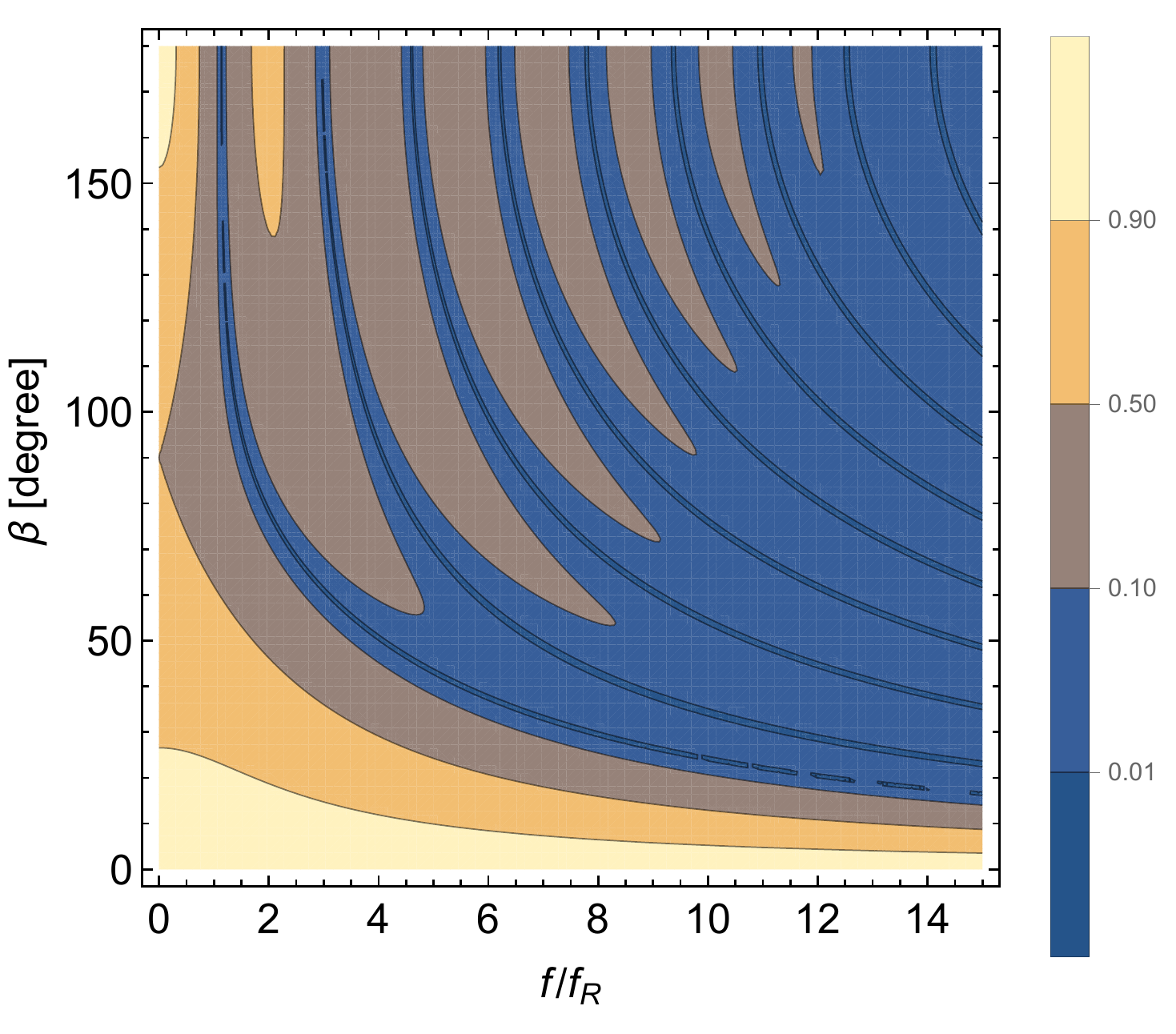}
\vspace{5mm}
\includegraphics[keepaspectratio, scale=0.6]{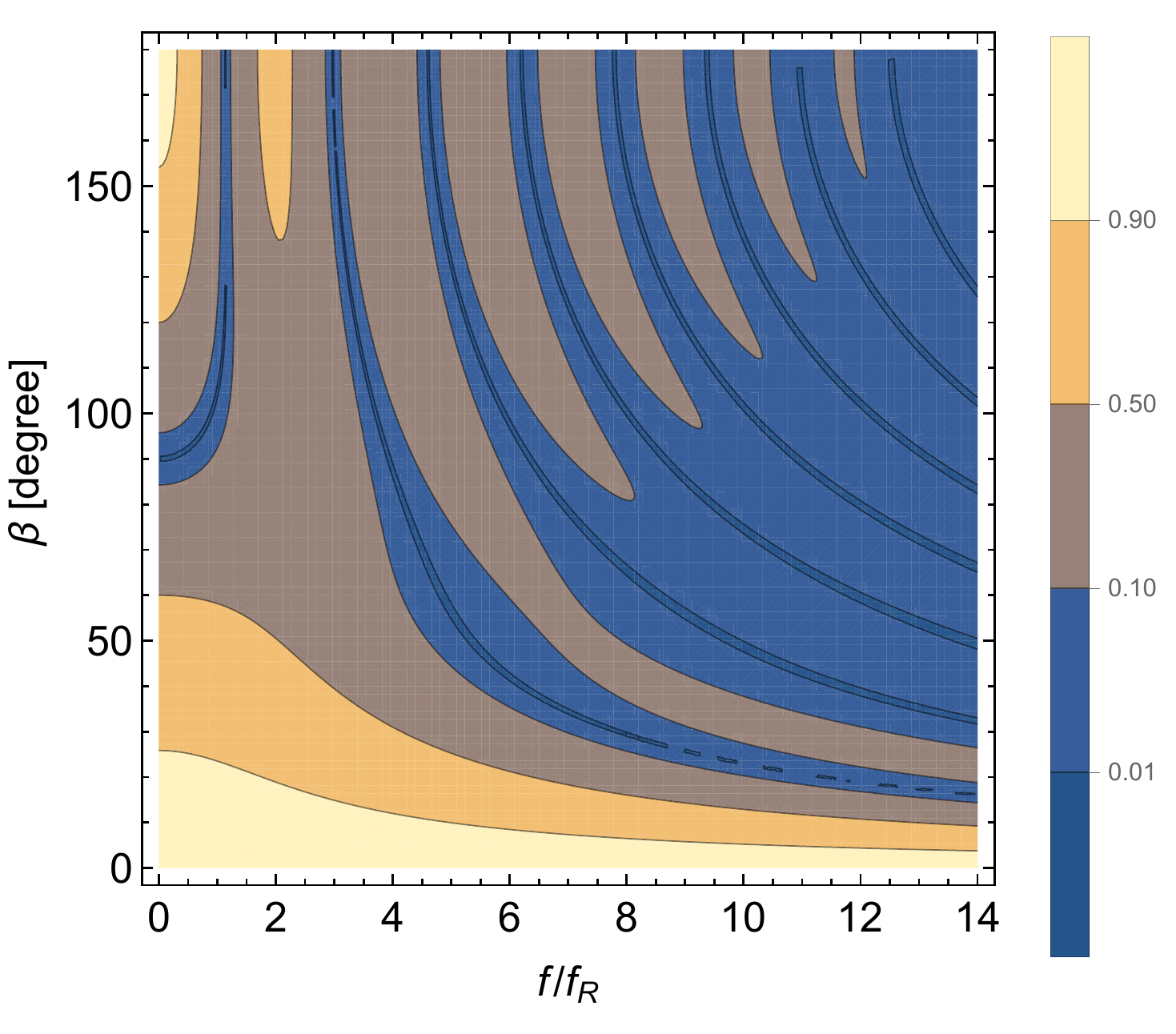}
\caption{Contour plots for the ORFs  $|\gamma_{AA'}|$ (upper)  and   $|\gamma_{EE'}|$ (lower). The interferometers $(A,E,A',E')$ are virtually aligned  by using the geodesic as in Fig. 3.  The parameter $\beta$  represents their  angular separation measured from the center of the contact sphere.  We   use the scaled  frequency $\eta=f/f_R$ with $f_R\equiv c/(2\pi R)$.  
}
\label{fig:2}
\end{figure}

\begin{figure}[t]
\centering
\includegraphics[keepaspectratio, scale=0.6]{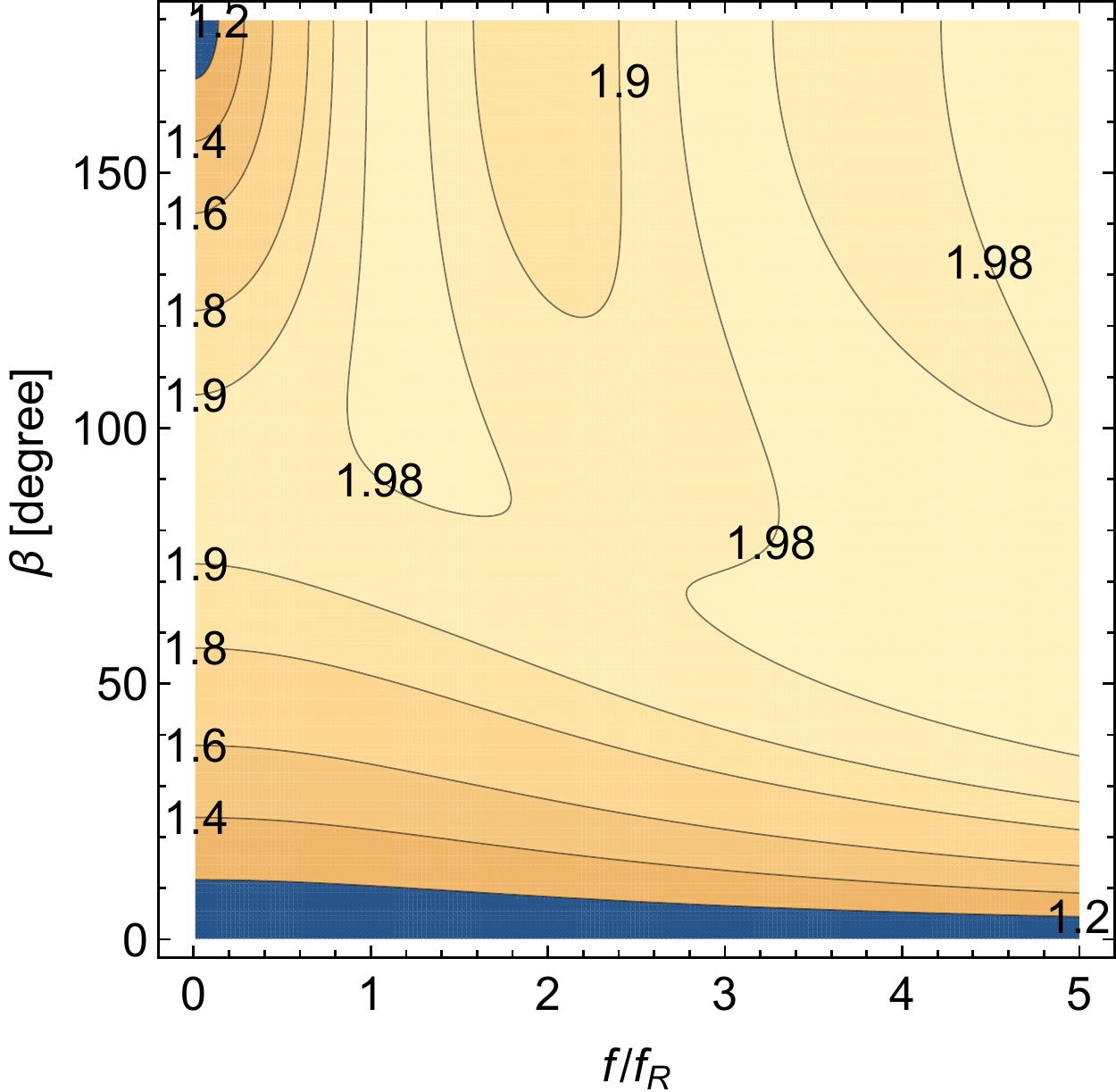}
\caption{Contour plot for the lower bound  $G_{\Delta \min}(\gamma_{AA'},\gamma_{EE'})$ defined in Eq. (\ref{min1}).  We have $f_R=$7.58Hz for ground based detectors and  $f_R=0.28$mHz for LISA-like detectors. 
}
\end{figure}

\subsection{Numerical Results}

Next we  concretely evaluate the two ORFs $\gamma_{AA'}$ and $\gamma_{EE'}$ for the aligned configuration (as in Fig. 3) on a sphere with radius $R$.   The spatial distance between the two units are given by 
$D=2R\sin(\beta/2)$  with the opening angle $\beta$ ($0^\circ \le \beta\le 180^\circ$) measured from the center of the contact sphere.  We define the characteristic frequency 
 $f_R\equiv {c}/{(2\pi R)}$, and introduce the rescaled one as
\begin{eqnarray}
\eta\equiv \frac{f}{f_R}.
\end{eqnarray}

Then, the two ORFs are expressed as \cite{Seto:2020zxw}
\begin{align}
\label{eq:14}
	\gamma_{AA'}(\eta,\beta) &= \Theta_1(y,\beta) - \Theta_2(y,\beta)~\\
	\gamma_{EE'} (\eta,\beta)&= \Theta_1(y,\beta) + \Theta_2(y,\beta)~
\end{align} 
with the variable
\begin{eqnarray}
y\equiv \frac{2\pi f D}c=2\sin\lmk\frac{\beta}2\rmk\eta.\label{defy}
\end{eqnarray}
The functions $\Theta_{1,2}$ are given by the spherical Bessel functions as
\begin{gather}
	\Theta_1(y,\beta) = \left(j_{0}(y) + \frac{5}{7}j_2(y) + \frac{3}{112} j_4(y)\right)\cos^4\left(\frac{\beta}{2}\right)\\
	\label{eq:17}
	\begin{aligned}
	\Theta_2(y,\beta) &= \left(-\frac{3}{8}j_0(y) + \frac{45}{56}j_2(y) - \frac{169}{896} j_4(y)\right)\\
	&+\left(\frac{1}{2}j_0(y) - \frac{5}{7}j_2(y) - \frac{27}{224} j_4(y)\right)\cos\beta\\
	&+\left(-\frac{1}{8}j_0(y) - \frac{5}{56}j_2(y) - \frac{3}{896} j_4(y)\right)\cos2\beta~.
	\end{aligned}
\end{gather}
In Fig. 5,  we present the contour plots for $|\gamma_{AA'}|$ and $|\gamma_{EE'}|$.  The dark blue lines roughly show the zero points of these functions.

Roughly speaking, because of the stronger phase coherence, the magnitudes of the ORFs become larger at lower frequency regime.   In concrete terms, we have $\lim_{y\to 0}j_n(y)=\delta_{n0}$ and the corresponding asymptotic profiles
\begin{eqnarray}
\gamma_{AA'}(0,\beta)=\frac{3+\cos(2\beta)}4,~~\gamma_{EE'}(0,\beta)=\cos(\beta)\label{g0}.
\end{eqnarray}
The absolute values of these expressions are minimum  both at $\beta=90^\circ$. 
In Fig. 5, we  also have $\gamma_{AA'}\to 1$ and $\gamma_{EE'}\to 1$ in the small angle limit  $\beta\to 0$ (resulting in $y\to 0$ from Eq. (\ref{defy})). 

 Given the ORFs, we can numerically evaluate the minimum value  $G_{\Delta \min}$ for the statistical gain.   In the next section, we discuss the preferred geometrical configuration for two units both  on the Earth and in space. As a preparation,  we provide the bound $G_{\Delta \min}$ as a function of the scaled frequency $\eta$ and the separation angle $\beta$
 \begin{eqnarray}
G_{\Delta\min}(\eta,\beta)=G_{\Delta\min}[(\gamma_{AA'}(\eta,\beta),\gamma_{EE'}(\eta,\beta)].
\end{eqnarray}

In Fig. 6, we present its contour plot.   We can observe significant potential {reduction of the gain} $G_{\Delta \min}\lsim 1.2$ at $\beta \lsim 10^\circ$, reflecting the strong correlation $\gamma_{AA'}\sim \gamma_{EE'}\sim 1$ shown in Fig. 5.  We also have the regime with  $G_{\Delta \min}\lsim 1.2$ at the upper left in this figure.  Actually, at $(\eta,\beta)=(0,180^\circ)$, we exactly obtain $G_{\Delta, \min}=1$  with $\gamma_{AA'}=1$ and $\gamma_{EE'}=-1$. 

In Fig. 6, on the vertical line $\eta=0$,   we have the peak value at $\beta=90^\circ$ 
\begin{eqnarray}
G_{\Delta\min}(\eta=0,\beta=90^\circ)=\frac{6+\sqrt3}{4}\simeq 1.933
\end{eqnarray}
directly from Eq. (\ref{g0}).

\section{application for future projects}

In this section, using Fig. 6, we discuss the statistical gains for networks composed by future triangular detectors. 

\begin{figure}[t]
\centering
\includegraphics[keepaspectratio, scale=0.45]{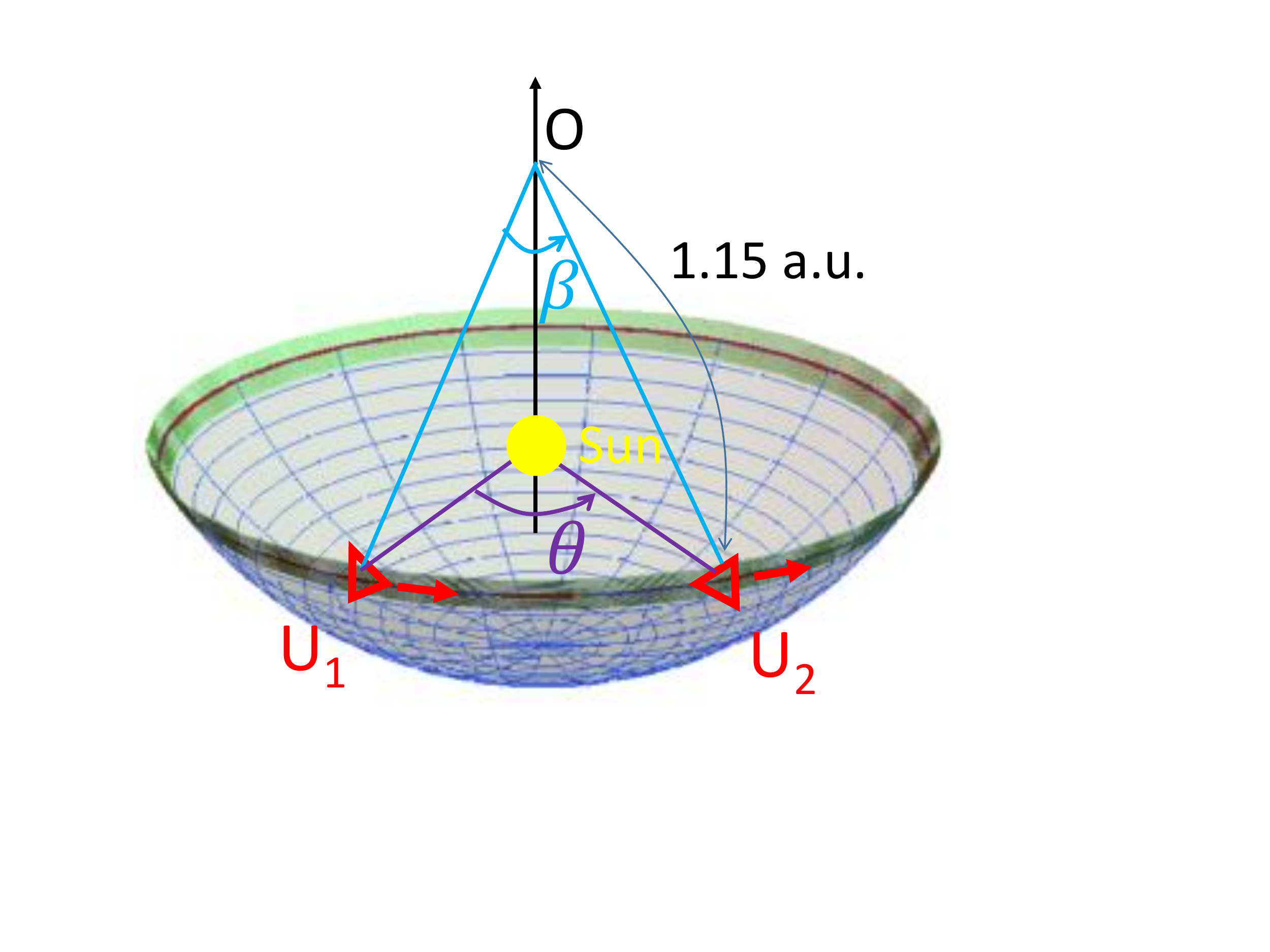}
\caption{Geometrical configuration of two LISA-like triangular units $U_1$ and $U_2$.   The two units move nearly on the ecliptic plane at the distance of  $\sim1$\,a.u. from the Sun with the phase difference $\theta$. The detector planes are inclined to the orbital (ecliptic) plane by $60^\circ$.  Their envelope (green belt) are tangential to a virtual sphere of  radius $R=1.15$\,a.u..   From the center O of the virtual sphere, the two units are separated by the angle $\beta$.  The relation between $\theta$ and $\beta$ is given in Eq. (\ref{bt}).
}
\end{figure}

\begin{figure}[t]
\centering
\includegraphics[keepaspectratio, scale=0.6]{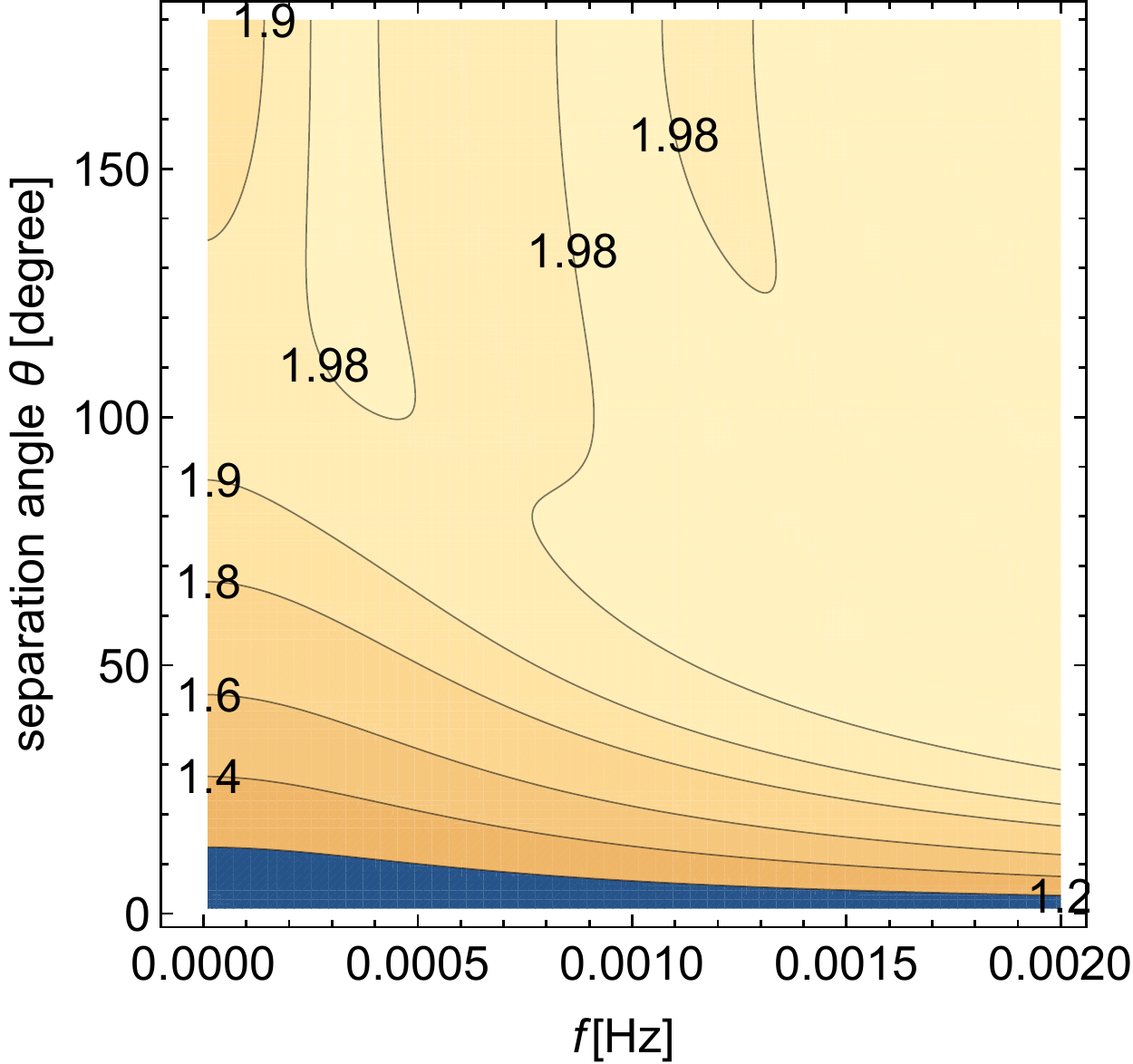}
\caption{Contour plot of the lower bound  $G_{\Delta \min}(\gamma_{AA'},\gamma_{EE'})$ for two LISA-like triangular units. In contrast to Fig. 6, the vertical axis is the orbital phase difference $\theta$, instead of $\beta$ (see  Fig. 7)  The proposed LISA-Taiji network has $\theta=40^\circ$.
}
\end{figure}

\subsection{Ground-Based Detectors}

First, we study two triangular detectors on the Earth, similar to ET.   The Earth has the characteristic frequency $f_R=7.58$Hz for its radius $R=6300$km.  The scaled frequency is given by $\eta=(f/{\rm 7.58 Hz})$.

At the low frequency regime, advanced LIGO (including its +-version) and advanced Virgo will have steep noise walls around 10Hz ($\eta\sim 1.3$). One of the major improvements of ET is to push down the noise wall down to $\sim 1$Hz ($\eta\sim 0.13$) and   open the new window at $\eta=0.1\sim 1$ \cite{Hild:2010id}.   

As an  example, let us assume that we have two triangular units separated by $\beta\sim 27^\circ$ (corresponding to the Hanford-Livingston distance).   Fig. 6 shows that the statistical gain  could be as small as $G_\Delta\sim 1.5$ in the new window $\eta=0.1\sim 1$. 
To guarantee a large statistical  gain $G_\Delta\gsim 1.9$ \add{(i.e. a small statistical loss $2-G_\Delta$)} in the frequency range, we need to set the separation in the range $70^\circ \lsim \beta \lsim 110^\circ$.

\subsection{LISA-Taiji Network}

LISA  is composed by three spacecrafts,  approximately forming a  regular triangular configuration.  It rotates around the Sun nearly on the ecliptic plane with the semi-major axis 1\,a.u..  As shown in  Fig. 7 with the green belt, the envelope of the  detector plane is inclined to the ecliptic plane by $\sim 60^\circ$.   We consider two LISA-like units that share the envelope of the detector plane,  separated by the angle $\theta$  corresponding  to the orbital phase difference.
For example, LISA and Taiji are planned to have the phase difference $\theta=40^\circ$ \cite{Seto:2020zxw,Omiya:2020fvw,Orlando:2020oko}(see also \cite{Wang:2021uih,Seto:2020mfd,Liang:2021bde,Wang:2021njt} for other configurations).

In fact, the envelope of the detector planes are tangential to a virtual sphere of radius  $R=2/\sqrt3=1.15$\,a.u..   The center O of the sphere is $1/\sqrt3$\,a.u. away from the Sun.  For the two units, the opening angle $\beta$ from the center O is written with the phase difference $\theta$ as \cite{Seto:2020zxw}
\begin{eqnarray}
\beta=\arccos \lkk \frac{(1+3\cos\theta)}4\rkk \equiv \beta(\theta)\label{bt}.
\end{eqnarray}
For example, we have correspondences $(\theta,\beta)=(0^\circ,0^\circ), (40^\circ,35.1^\circ), (109.4^\circ,90^\circ)$ and $(180^\circ,120^\circ)$.

Now, we can apply Fig. 6  for two LISA-like units.    For the virtual sphere with $R=1.15$\,a.u., the characteristic frequency is $f_R=0.28$\,mHz with the scaled one $\eta=(f/{\rm 0.28 mHz})$. 
In Fig. 8, we show the contour plot generated from Fig. 6.
This figure  shows that, with the   current design $\theta=40^\circ$ for the LISA-Taiji network, we might have a \add{relatively small gain} $G_\Delta\sim 1.6$ at $f\lsim 0.5$\,mHz ($\eta\lsim 1.8$). To realize $G_\Delta> 1.9$ at the whole frequency regime, we need to take the orbital phase difference  at $89^\circ < \theta< 153^\circ$ (corresponding to $75^\circ<\beta<105^\circ$).  At the low frequency regime,   the optimal choice  is $\theta=109.4^\circ$ ($\beta=90^\circ$) .

\subsection{Galactic Confusion Noise}

Next, we discuss the reduction of the statistical gain caused by the Galactic confusion noise    made from unresolved Galactic binaries in the LISA band. In Fig. 9, we present the instrumental noise $N_D(f)$ of LISA  and the estimated  Galactic confusion noise spectrum $N_B(f)$ \cite{Cornish:2018dyw}. 

In reality, the Galactic confusion background is anisotropic (see e.g. \cite{Giampieri,Ungarelli:2001xu,Seto:2004ji,Edlund:2005ye,Littenberg:2020bxy}).  Since the orientations of the triangle units change with time, their confusion noise spectra will be  time dependent. Therefore, the  spectrum $N_B(f)$ in Fig. 9 should be regarded as a time-averaged one. 

In principle, we can develop a formulation for evaluating the statistical gain (as an extension of Secs. II and III) including the anisotropies of the Galactic background.  However, we can no longer use the geometrical symmetries associated with   isotropic backgrounds, and  the intermediate calculations become  much more complicated. For example, we cannot make the $2\times2$ block diagonalization with the alignment  based on the geodesic (see Fig. 3).  Furthermore, some of the correlation coefficients (e.g. $\gamma_{AA'}$) would be  complex numbers, due to the odd multipoles of the incoming background waves. 

\begin{figure}[t]
\centering
\includegraphics[keepaspectratio, scale=0.5]{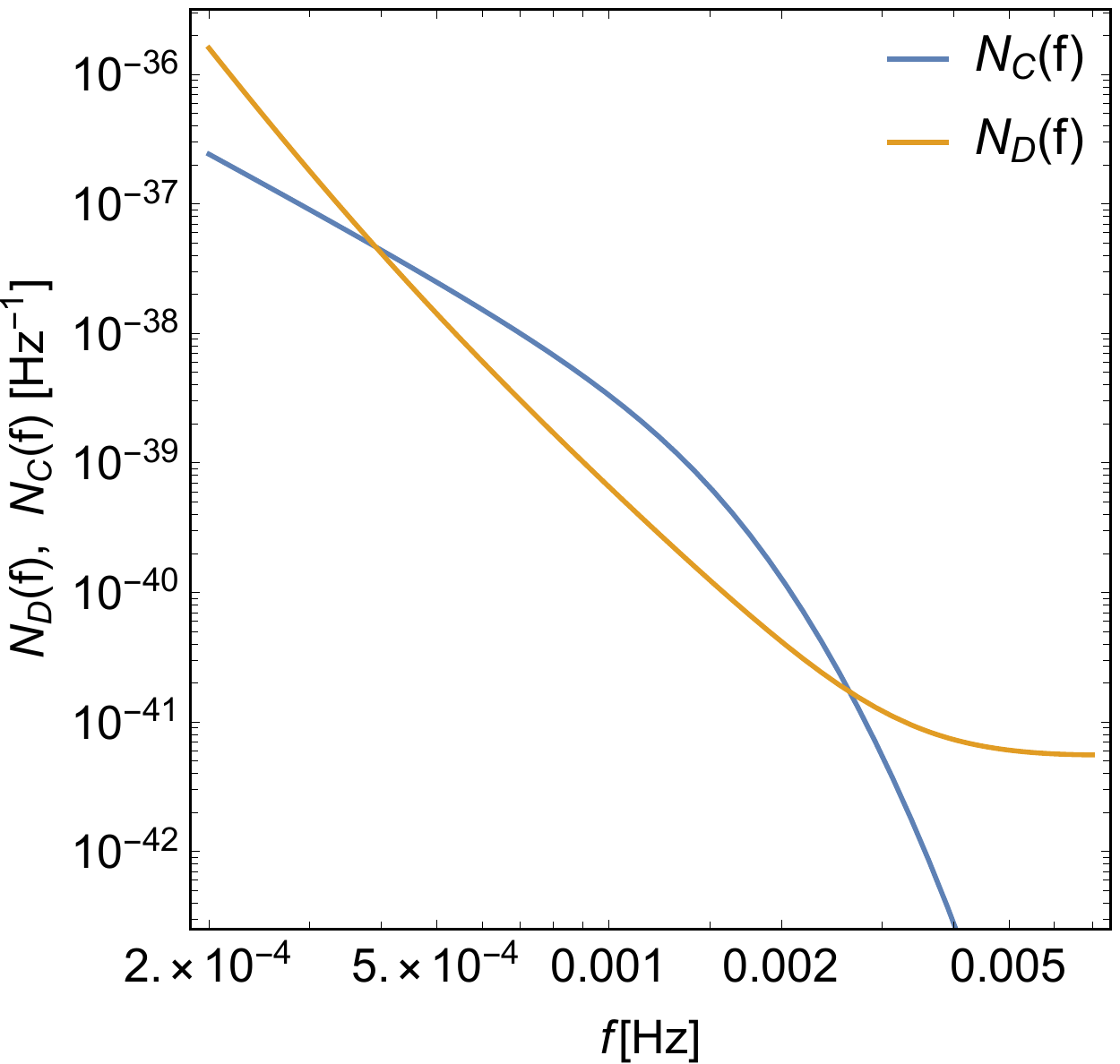}
\caption{The detector noise  spectrum  $N_{D}(f)$ of LISA and the estimated Galactic confusion noise spectrum $N_{C}(f)$  after 4yr integration \cite{Cornish:2018dyw}.   These spectra  are presented for the single L-shaped  interferometers  (corresponding to the $A$ or $E$ modes)  without the angular averaging.  The relative noise strength is given by $K(f)= N_{C}(f)/N_{D}(f) $. 
}
\end{figure}

\begin{figure}[t]
\centering
\includegraphics[keepaspectratio, scale=0.6]{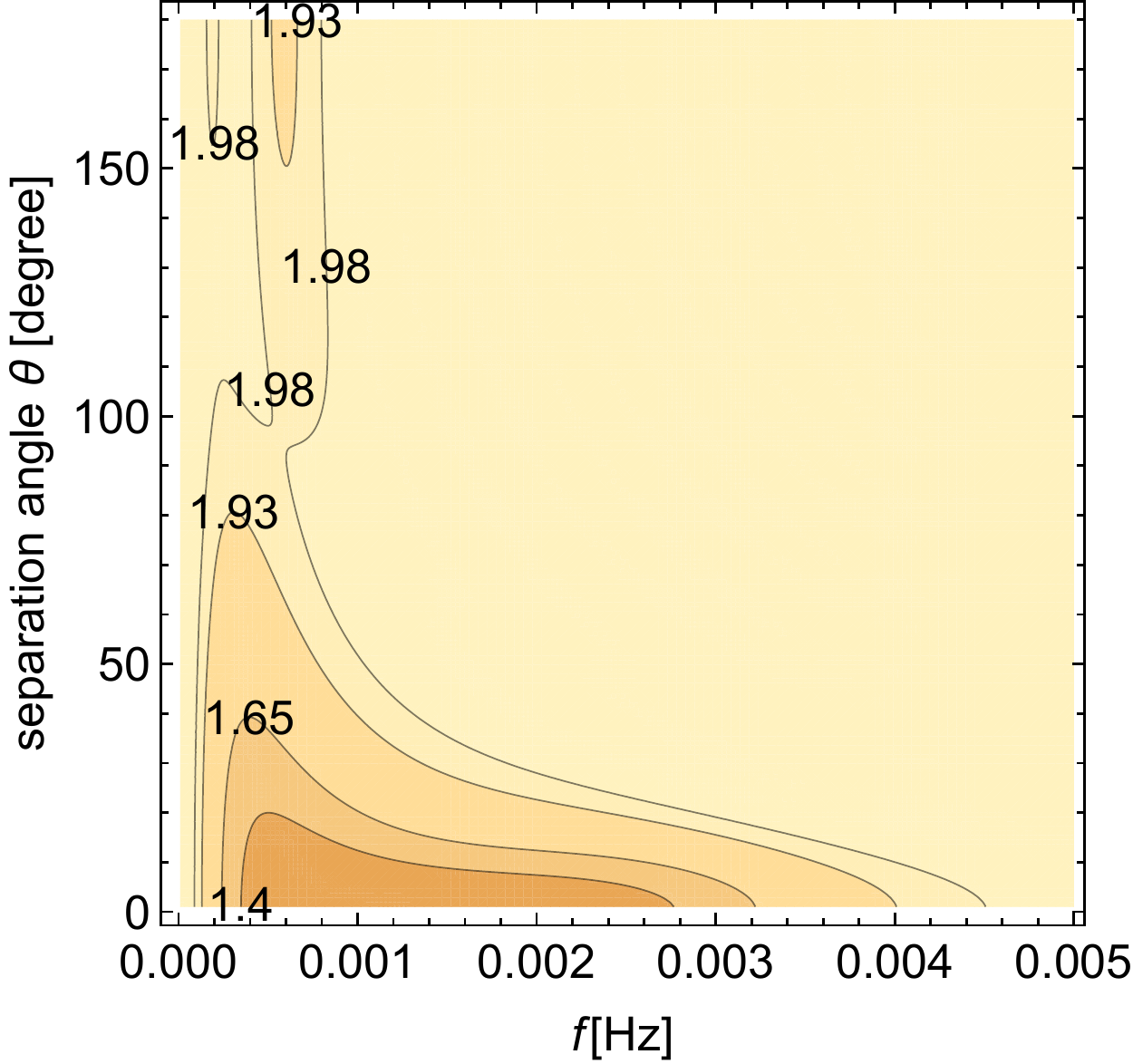}
\caption{The statistical gain $ G_{\Delta}(K(f),\gamma_{AA'}(f), \gamma_{EE'}(f))$ for two LISA-like units.  The vertical axis represents the orbital phase difference $\theta$.  
}
\end{figure}

Here, ignoring the anisotropies, we simply apply our expression $G_\Delta(K,\gamma_{AA'},\gamma_{EE'})$ in Eq. (32) for approximately evaluating the statistical gain affected by the Galactic confusion noise.  We use the ORFs  $\gamma_{AA'}$ and 
 $\gamma_{EE'}$ defined for isotropic backgrounds and also plug in  the time averaged ratio 
$K(f)=N_B(f)/N_D(f)$ shown in Fig. 9.

\add{ For an anisotropic background, the degree of the noise correlation (corresponding to the ORFs for an isotropic case) depends on the overall  orientation of the network.  Therefore, it will not be  straightforward to tell  whether the present approximation overestimates or underestimates the actual statistical gain.   }

We expect that our simplified treatment would be a convenient approximation, roughly taking into account the effective averaging induced by the rotation of the detectors (see Fig. 7).  It should be also noticed that, unlike the minimum value  $G_{\Delta \min} (\gamma_{AA'},\gamma_{EE'})$ in Eq. (\ref{min1}), we now keep the explicit $K$-dependence of the original expression  $G_{\Delta} (K,\gamma_{AA'},\gamma_{EE'})$ for the  statistical gain.  This allows us to see the cooperation of the noise ratio $K$ and the ORFs  $(\gamma_{AA'},\gamma_{EE'})$  for the statistical gain.

In Fig. 10,  we show our numerical results.   We can observe a relatively small gain (e.g. $G_\Delta\lsim 1.4$) only around the frequency regime $0.2{\rm mHz}\lsim f\lsim 3$mHz where the magnitude of the confusion noise becomes $K\gsim 1$. 
For  the angle $\theta=40^\circ$ currently designed for the LISA-Taiji network, we could have  $G_\Delta\sim 1.65$ around $f\sim 0.4$mHz.  If we take $80^\circ<\theta<180^\circ$, the gain could be         $G_\Delta\gsim 1.93$ in all the frequency range.

So far, we have assumed that  the confusion noise level $N_B(f)$ is independent of the orbital configuration of the detector network (not only ignoring its dependence on the numbers of the units). But, in Fig. 10,  the subtraction of the Galactic binaries is likely to work more efficiently at the regions with higher $G_\Delta$.   Then, the residual noise $N_B(f)$ itself would become  smaller at the corresponding regions.    Therefore, the actual contrast of  the ratio $\sigma_{2U}/\sigma_{1U}$ would be  somewhat larger than Fig. 10.

\section{Summary and discussion}

In this paper, we studied performance of gravitational wave networks under the existence of coherent background noises.  We introduced the effective number of detectors to characterize the statistical {gain} of the angular averaged sensitivity for  a detector  network. 

We first examined the basic model for two L-shaped interferometers and derived the expression $G_2(K,\gamma)$ with the noise ratio $K$ and the ORF $\gamma$.  We discussed the overall properties of this expression, including its minimum value and the asymptotic profiles. 

We then examined  the effective number $G_\Delta$ for  networks composed by two triangular detectors tangential to a sphere (see Eq. (\ref{gd})).  By using the symmetries of the systems, we could handle the problem  as a straightforward extension of the basic model.     The  expression   $G_\Delta$  can be applied not only to  ground  based detector networks but also to space detectors such as the LISA-Taiji network.  The related expression (\ref{gd2}) would be useful for a weak background level $K\ll 1$.  

The magnitude of the noise ratio $K$ cannot be securely measured beforehand  for  an essentially new frequency window. Therefore, as the worst case,   we calculated the minimum value of the expression $G_\Delta$, and discussed the preferable network geometry to suppress the potential reduction  of the statistical gain.  We can guarantee { large gains} $G_\Delta\gsim 1.9$ for the angular separation in the range $70^\circ <\beta <105^\circ$, corresponding to the orbital phase difference  $89^\circ <\theta <153^\circ$ for space detectors (see Figs. 6 and  8).

Finally, under the approximation that the Galactic confusion background is isotropic,  we examined its impacts on the two LISA-like units, by plugging in the estimated  ratio $K$.  
With the currently proposed value $\theta=40^\circ$ for the LISA-Taiji network, we have $G_\Delta \sim 1.65$ around 0.4mHz.  By taking the orbital phase difference  $\theta>80^\circ$, we can \add{increase the statistical gains} to $G_\Delta \gsim 1.93$ in all the frequency range.

\add{ For suppressing  the background noise correlation, our basic strategy was to   reduce the absolute values of the ORFs.  But, this is the opposite direction for enhancing the sensitivity to  backgrounds by correlation analysis.  It might be worth considering to take a balance between the two requirements. }

In this paper, we have  focused on the angular averaged sensitivity, as one of the fundamental quantities to characterize detector networks. But it would be interesting to examine impacts of the background noise correlation on other measures.   
Considering the active studies on the space detector networks, it would be also meaningful to extend  the present study to include anisotropies of the Galactic confusion background and make detailed studies on the preferable network geometry.

\begin{acknowledgments}
The author would like to thank H. Omiya for useful
conversations.
 This work is supported by JSPS Kakenhi Grant-in-Aid for Scientific Research
 (Nos. 17H06358 and 19K03870).
\end{acknowledgments}





\end{document}